\documentclass[prd,twocolumn,showpacs,floatfix,amsmath,nofootinbib,amssymb,floatfix]{revtex4}
\usepackage{graphicx,color,dcolumn,booktabs,bm}
\usepackage{longtable,lscape}
\usepackage{txfonts}
\usepackage{overpic}
\usepackage{amssymb}
\usepackage{indentfirst}
\usepackage{feynmf}   
\usepackage{slashed}  
\usepackage{cases}
\usepackage{color}
\usepackage{multirow}
\usepackage{epstopdf}
\usepackage{graphicx,color,dcolumn,booktabs,bm}
\usepackage[colorlinks, citecolor=blue,anchorcolor=red,menucolor=red, linkcolor=red,filecolor=red,runcolor=red,urlcolor=blue,frenchlinks=red]{hyperref}

\graphicspath{{Figures/}} %

\begin{document}

\title{$D_{s1}^*(2860)$ and $D_{s3}^*(2860)$: Candidates for $1D$ charmed-strange mesons}
\author{Qin-Tao Song$^{1,2,6}$}
\author{Dian-Yong Chen$^{1,2}$\footnote{Corresponding author}
}\email{chendy@impcas.ac.cn}
\author{Xiang Liu$^{2,3}$\footnote{Corresponding author}}\email{xiangliu@lzu.edu.cn}
\author{Takayuki Matsuki$^{4,5}$}\email{matsuki@tokyo-kasei.ac.jp}
\affiliation{$^1$Nuclear Theory Group, Institute of Modern Physics, Chinese Academy of Sciences, Lanzhou 730000, China\\
$^2$Research Center for Hadron and CSR Physics,
Lanzhou University $\&$ Institute of Modern Physics of CAS,
Lanzhou 730000, China\\
$^3$School of Physical Science and Technology, Lanzhou University,
Lanzhou 730000, China\\
$^4$Tokyo Kasei University, 1-18-1 Kaga, Itabashi, Tokyo 173-8602, Japan\\
$^5$Theoretical Research Division, Nishina Center, RIKEN, Saitama 351-0198, Japan\\
$^6$ University of Chinese Academy of Sciences, Beijing 100049, China}

\begin{abstract}
Newly observed two charmed-strange resonances, $D_{s1}^*(2860)$ and $D_{s3}^*(2860)$, are investigated by calculating their Okubo-Zweig-Iizuka allowed strong decays, which shows that they are suitable candidates for the $1^3D_1$ and $1^3D_3$ states in the charmed-strange meson family. Our study also predicts other main decay modes of $D_{s1}^*(2860)$ and $D_{s3}^*(2860)$, which can be accessible at the future experiment. In addition, the decay behaviors of the spin partners of $D_{s1}^*(2860)$ and $D_{s3}^*(2860)$, i.e., $1D(2^-)$ and $1D^\prime(2^-)$, are predicted in this work, which are still missing at present. Experimental search for the missing $1D(2^-)$ and $1D^\prime(2^-)$ charmed-strange mesons is an intriguing and challenging task for further experiment.
\end{abstract}

 \maketitle

\section{introduction}\label{sec1}

Very recently the LHCb Collaboration has released a new observation of
an excess around 2.86 GeV in the $\bar{D}^0K^-$ invariant mass
spectrum of $B_s^0\to \bar{D}^0K^-\pi^+$, which can be an
admixture of spin-1 and spin-3 resonances corresponding to
$D_{s1}^*(2860)$ and $D_{s3}^*(2860)$
\cite{Aaij:2014xza,Aaij:2014baa}, respectively. As indicated by
LHCb \cite{Aaij:2014xza,Aaij:2014baa}, it is the first time to
identify a spin-3 resonance. In addition, $D_{s2}^*(2573)$ also
appears in the the $\bar{D}^0K^-$ invariant mass spectrum.

Before this observation, a charmed-strange state
$D_{sJ}(2860)$ was reported by BaBar in the $DK$ channel
\cite{Aubert:2006mh}, where the mass and width are
$m=2856.6\pm1.5\pm5.0$ MeV and $\Gamma=47\pm7\pm10$ MeV
\cite{Aubert:2006mh}, respectively, which was later confirmed by BaBar in
the $D^*K$ channel \cite{Aubert:2009ah}. The $D_{sJ}(2860)$
has stimulated extensive discussions on its underlying structure. In
Ref.~\cite{Zhang:2006yj}, $D_{sJ}(2860)$ is suggested as a $1^3D_3$
$c\bar{s}$ meson. This explanation was also supported by study
of the effective Lagrangian approach \cite{Colangelo:2006rq,Colangelo:2012xi}, the Regge phenomenology \cite{Li:2007px}, constituent quark model
\cite{Zhong:2009sk} and mass loaded flux tube model
\cite{Chen:2009zt}. The ratio $\Gamma(D_{sJ}(2860)\to D^*K)/\Gamma(D_{sJ}(2860)\to DK)$
was calculated as 0.36 \cite{Colangelo:2007ds} by the effective Lagrangian
method. However, the calculation
by the QPC model shows that such a ratio is about 0.8
\cite{Li:2009qu}, which is close to the experimental value
$1.10\pm0.15\pm0.19$ \cite{Aubert:2009ah}. Thus, $J^P=3^-$
assignment to $D_{sJ}(2860)$ is a possible explanation. In addition,
$D_{sJ}(2860)$ as a mixture of charmed-strange states was given in
Refs.~\cite{Li:2009qu,Zhong:2008kd,Zhong:2009sk}. $D_{sJ}(2860)$ could be a
partner of $D_{s1}(2710)$, where both $D_{sJ}(2860)$ and
$D_{s1}(2710)$ are a mixture of $2^3S_1$ and $1^3D_1$ $c\bar{s}$
states. By introducing such a mixing mechanism, the obtained ratio of
$D^*K/DK$ for $D_{sJ}(2860)$ and $D_{s1}(2710)$ \cite{Li:2009qu} is
consistent with the experimental data \cite{Aubert:2009ah}.
Reference \cite{vanBeveren:2009jq} indicates that there exist two
overlapping resonances (radially excited $J^P=0^+$ and $J^P=2^+$
$c\bar{s}$ states) at 2.86 GeV. Besides the above explanations under
the conventional charmed-strange meson framework, $D_{sJ}(2860)$ was
explained as a multiquark exotic state \cite{Vijande:2008zn}.
{$D_{sJ}(2860)$ as a $J^P=0^+$ charmed-strange meson was suggested in Ref. \cite{Zhang:2006yj}. However, this scalar charmed-strange meson cannot decay into $D^*K$ \cite{Zhang:2006yj}, which contradicts the BaBar's observation of $D_{sJ}(2860)$ in its $D^*K$ decay channel \cite{Aubert:2006mh}. After the observation of $D_{sJ}(2860)$, $D_{sJ}(3040)$ was reported by BaBar in the $D^*K$ channel \cite{Aubert:2009ah}, which can be explained as the first radial excitation of $D_{s1}(2460)$ with $J^P=1^+$ \cite{Sun:2009tg}. In addition, the decay behaviors of other missing $2P$ charmed-strange mesons in experiment were given in Ref. \cite{Sun:2009tg}. }

Briefly reviewing the research status of $D_{sJ}(2860)$, we
notice that more theoretical and experimental efforts
are still necessary to clarify the properties of
$D_{sJ}(2860)$. It is obvious that the recent precise
measurement of LHCb \cite{Aaij:2014xza,Aaij:2014baa} provides us a good
chance to identify higher radial excitations in the charmed-strange
meson family.

The resonance parameters of the newly observed $D_{s1}^*(2860)$ and
$D_{s3}^*(2860)$ by LHCb include \cite{Aaij:2014xza,Aaij:2014baa}:
\begin{eqnarray}
m_{D_{s1}^*(2860)}&=&2859\pm12\pm6\pm23\,\, {\mathrm{MeV}},\\
\Gamma_{D_{s1}^*(2860)}&=&159\pm23\pm27\pm72\,\, {\mathrm{MeV}},\\
m_{D_{s3}^*(2860)}&=&2860.5\pm2.6\pm2.5\pm6.0\,\,{\mathrm{MeV}},\\
\Gamma_{D_{s3}^*(2860)}&=&53\pm7\pm4\pm6\,\, {\mathrm{MeV}},
\end{eqnarray}
where the errors are due to statistical one,
experimentally systematic effects and model
variations \cite{Aaij:2014xza,Aaij:2014baa}, respectively.

At present, there are good candidates for the $1S$ and $1P$ states
in the charmed-strange meson family (see Particle Data Group for
more details \cite{Beringer:1900zz}). Thus, two newly observed
$D_{s1}^*(2860)$ and $D_{s3}^*(2860)$ can be categorized into the
$1D$ charmed-strange states when considering their
spin quantum numbers and masses. Before observation
of these two resonances, there were several theoretical predictions
\cite{Godfrey:1985xj, Matsuki:2007zza, Di Pierro:2001uu,
Ebert:2009ua} of the mass spectrum of the $1D$
charmed-strange meson family, which are collected in Table
\ref{prediction}. Comparing the experimental data of
$D_{s1}^*(2860)$ and $D_{s3}^*(2860)$ with the theoretical results,
we notice that the masses of $D_{s1}^*(2860)$ and $D_{s3}^*(2860)$
are comparable with the corresponding theoretical predictions, which
further supports that it is reasonable to assign $D_{s1}^*(2860)$ and
$D_{s3}^*(2860)$ as the $1D$ states of the charmed-strange meson family

\renewcommand{\arraystretch}{1.5}
\begin{table}[htbp]
\caption{Theoretical predictions for charmed-strange meson spectrum
and comparison with the experimental data. \label{prediction}}
\begin{tabular}{lccccc}
\toprule[1pt]%
$J^P(^{2s+1}L_J)$ & Expt. \cite{Beringer:1900zz} & GI \cite{Godfrey:1985xj} &
MMS \cite{Matsuki:2007zza}
& PE \cite{Di Pierro:2001uu}  & EFG \cite{Ebert:2009ua} \\
\hline%
 $0^-(^1S_0)$ & 1968 & 1979 & 1967 & 1965 & 1969 \\
 $1^-(^3S_1)$ & 2112 & 2129 & 2110 & 2113 & 2111 \\
\hline%
 $0^+(^3P_0)$ & 2318 & 2484 & 2325 & 2487 & 2509 \\
$1^+("^1P_1")$& 2460 & 2459 & 2467 & 2535 & 2536 \\
$1^+("^3P_1")$& 2536 & 2556 & 2525 & 2605 & 2574 \\
 $2^+(^3P_2)$ & 2573 \cite{Aaij:2014xza, Aaij:2014baa} & 2592 & 2568 & 2581 & 2571 \\
\hline%
 $1^-(^3D_1)$ & 2859 \cite{Aaij:2014xza, Aaij:2014baa} & 2899 & 2817 & 2913 & 2913 \\
$2^-("^1D_2")$&  --  & 2900 &  --  & 2900 & 2931 \\
$2^-("^3D_2")$&  --  & 2926 & 2856 & 2953 & 2961 \\
 $3^-(^3D_3)$ & 2860 \cite{Aaij:2014xza, Aaij:2014baa} & 2917 &  --  & 2925 & 2871 \\
\bottomrule[1pt]%
\end{tabular}
\end{table}

Although both the mass spectrum analysis and the measurement of the
spin quantum number support $D_{s1}^*(2860)$ and $D_{s3}^*(2860)$ as
the $1D$ states, we still need to carry out a further test of this
assignment through study of their decay behaviors. This study can
provide more detailed information on the partial decay widths,
which is valuable for future experimental investigation of
$D_{s1}^*(2860)$ and $D_{s3}^*(2860)$. In addition, there exist four
$1D$ states in the charmed-strange meson family. At present, the
spin partners of $D_{s1}^*(2860)$ and $D_{s3}^*(2860)$ are still
missing in experiment.  Thus, we will also predict the
properties of two missing $1D$ states in this work.

This paper is organized as follows. In Section \ref{sec2}, after some introduction we illustrate the study of decay behaviors of $D_{s1}(2860)$ and $D_{s3}(2860)$. In Sec. \ref{sec3}, the paper ends with the discussion and conclusions.

\section{Decay behavior of $D_{s1}(2860)$ and $D_{s3}(2860)$}\label{sec2}

Among all properties of these $1D$ states, their two-body
Okubo-Zweig-Iizuka (OZI)-allowed strong decays are the
most important and typical properties. Hence, in the following we
mainly focus on the study of their OZI-allowed strong decays. For
the $1D$ states studied in this work, their allowed decay
channels are listed in Table \ref{table:channel}. Among four $1D$
states in the charmed-strange meson family, there are two $J^P=2^-$
states, which is a mixture of $1^1D_2$ and $1^3D_2$ states,
i.e.,
\begin{eqnarray}
\left(\begin{array}{c}1D(2^-)
\\1D^\prime(2^-)\end{array}\right)=\left(\begin{array}{cc}\cos\theta_{1D}
& \sin\theta_{1D} \\-\sin\theta_{1D} &
\cos\theta_{1D}\end{array}\right) \left(\begin{array}{c}1^3D_2
\\1^1D_2\end{array}\right),
\end{eqnarray}
where $\theta_{1D}$ is a mixing angle. { In the heavy quark limit, a general mixing angle between  $^3L_L$ and $^1L_L$ is $\theta_{L}=\arctan(\sqrt{L/(L+1)})$, which indicates $\theta_{1D}=39^\circ$ \cite{Matsuki:2010zy}.}

\begin{table}[htb]
\caption{The two-body OZI-allowed decay modes of $1D$
charmed-strange mesons. Here, we use symbols, $\circ$ and --, to mark the
OZI-allowed and -forbidden decay modes, respectively. $D_{s1}^*(2860)$
and  $D_{s3}^*(2860)$ are $1^3D_1$ and $1^3D_3$ states,
respectively. \label{table:channel} } \centering
\begin{tabular}{l c c c c c c c c c c c c c}
\toprule[1pt]%
\multirow{2}{*}{Channels}  & \multirow{2}{*}{$D^*_{s1}(2860)$} &
$1D(2^-)$  &  \multirow{2}{*}{$D_{s3}^*(2860)$}\\
                      &     &       $1D'(2^-)$  &\\
 \hline%
$D K$                 &$\circ$  &  --     &$\circ$  \\
$D^{*}K$              &$\circ$  &$\circ$  &$\circ$  \\
$D_s\eta$             &$\circ$  &  --     &$\circ$  \\
$D_{s}^*\eta$         &$\circ$  &$\circ$  &$\circ$  \\
$DK^*$                &$\circ$  &$\circ$  &$\circ$  \\
$D_0^*(2400)K$        &  --     &$\circ$  &  --     \\
$D_{s0}^*(2317)\eta$  &  --     &$\circ$  &  --     \\
\bottomrule[1pt]%
\end {tabular}\\
\label{table:decay}
\end{table}

In the following, we apply the quark pair creation (QPC) model
\cite{Micu:1968mk,Le Yaouanc:1972ae, LeYaouanc:1988fx,
vanBeveren:1979bd, vanBeveren:1982qb, Bonnaz:2001aj, roberts} to
describe the OZI allowed two-body strong decays shown in Table
\ref{table:channel}, where the QPC model was extensively adopted to
study the strong decay of hadrons \cite{Zhang:2006yj, Liu:2009fe,
Sun:2009tg, Sun:2010pg, Yu:2011ta, Wang:2012wa, Ye:2012gu,
He:2013ttg, Sun:2013qca, Sun:2014wea, Pang:2014laa, He:2014xna}. In the QPC model, the quark-antiquark pair is created in QCD vacuum with vacuum quantum number $J^{PC}=0^{++}$. For a decay process, i.e., an initial observed meson $A$ decaying into two observed mesons $B$ and $C$, the process can be expressed as
\begin{eqnarray}
\langle BC|T|A \rangle = \delta ^3(\mathbf{P}_B+\mathbf{P}_C)\mathcal{M}^{{M}_{J_{A}}M_{J_{B}}M_{J_{C}}},\label{hh1}
\end{eqnarray}
where $\mathbf{P}_B(\mathbf{P}_C)$ is the three-momentum of the final meson $B(C)$ in the rest frame of $A$. $M_{J_{i}} (i=A,B,C)$ denotes the orbital magnetic momentum. Additionally, $\mathcal{M}^{{M}_{J_{A}}M_{J_{B}}M_{J_{C}}}$ is the helicity amplitude. The transition operator $T$ in Eq. (\ref{hh1}) is written as \cite{Micu:1968mk,Le
Yaouanc:1972ae,LeYaouanc:1988fx,vanBeveren:1979bd,
vanBeveren:1982qb,Bonnaz:2001aj,roberts}
\begin{eqnarray}
T& = &-3\gamma \sum_{m}\langle 1m;1-m|00\rangle\int d \mathbf{p}_3d\mathbf{p}_4\delta ^3 (\mathbf{p}_3+\mathbf{p}_4) \nonumber \\
 && \times \mathcal{Y}_{1m}\left(\frac{\textbf{p}_3-\mathbf{p}_4}{2}\right)\chi _{1,-m}^{34}\phi _{0}^{34}
\omega_{0}^{34}b_{3i}^{\dag}(\mathbf{p}_3)d_{4j}^{\dag}(\mathbf{p}_4),
\end{eqnarray}
which is introduced in a phenomenological way to reflect the property of quark-antiquark (denoted by indices 3 and 4) created from vacuum. $\mathcal{Y}_{lm}(\mathbf{p})=|\mathbf{p}|Y_{lm}(\mathbf{p}) $ is the solid harmonic. $\chi$, $\phi$, and $\omega$ are the general description of the spin, flavor, and color wave functions, respectively.

By the Jacob-Wick formula \cite{Jacob:1959at}, the decay amplitude reads as
\begin{eqnarray}
\mathcal{M}^{JL}(\mathbf{P})&=&\frac{\sqrt{2L+1}}{2J_A+1}\sum_{M_{J_B}M_{J_C}}\langle L0;JM_{J_A}|J_AM_{J_A}\rangle \nonumber \\
&&\times \langle J_BM_{J_B};J_CM_{J_C}|{J_A}M_{J_A}\rangle \mathcal{M}^{M_{J_{A}}M_{J_B}M_{J_C}}.
\end{eqnarray}
Finally, the general decay width is
\begin{eqnarray}
\Gamma&=&\pi ^2\frac{|\mathbf{P}_B|}{m_A^2}\sum_{J,L}|\mathcal{M}^{JL}(\mathbf{P})|^2,
\end{eqnarray}
where $m_{A}$ is the mass of the initial state $A$.
In the following, the helicity
amplitudes $\mathcal{M}^{M_{J_A}M_{J_B}M_{J_C}}$ of
the OZI-allowed strong decay channels in Table \ref{table:channel}
are calculated, which is the main task of the whole calculation. { Here, we adopt the simple harmonic oscillator (SHO) wave function $\Psi_{n,\ell
m}(\mathbf{k})$, where the value of a parameter $R$ appearing in the SHO wave function can be obtained such that it reproduces the realistic root mean square (rms) radius, which can be calculated by the relativistic quark model \cite{Close:2005se} with a Coulomb plus linear confining potential as well as a hyperfine interaction term.}
In Table \ref{table:Rvalue}, we list the $R$ values adopted
in our calculation. {The strength of $q\bar{q}$ is taken as
$\gamma=6.3$ \cite{Sun:2009tg} while the strength of $s\bar{s}$ satisfies
$\gamma_s=\gamma/\sqrt{3}$. We need to specify our $\gamma$ value adopted in the present work which is $\sqrt{96\pi}$ times larger than that adopted by other groups \cite{Godfrey:1986wj,Close:2005se}, where the $\gamma$ value
as an overall factor can be obtained by fitting the experimental data (see Ref. \cite{Blundell:1996as} for more details of how to get the $\gamma$ value).}
 In addition, the
constituent quark masses for charm, up/down, and strange quarks are
1.60 GeV, 0.33 GeV, and 0.55 GeV, respectively \cite{Close:2005se}.
 \begin{table}[ht]
\caption{The $R$ values (in units of GeV$^{-1}$) \cite{Close:2005se}
and masses (in units of MeV) \cite{Beringer:1900zz} of the mesons involved in
present calculation. \label{table:Rvalue} } \centering
\begin{tabular}{ccccccccc}
\toprule[1pt]%
States& $R$ & mass&States& $R$ &mass &States& $R$ &mass \\
\midrule[1pt]
$D$   & 2.33 & 1867 &$D^\ast$  & 2.70   & 2008 &$D_0(2400)$    &  3.13 & 2318\\
$D_s$ & 1.92 & 1968 &$D_s^\ast$& 2.22   & 2112 &$D_{s0}(2317)$ &  2.70 & 2318\\
$K$   & 2.17 & 494  & $K^\ast$ & 3.13   & 896  &   $\eta$      &  2.12 & 548 \\
\bottomrule[1pt]
\end {tabular}
\end{table}

\begin{figure*}[htbp]
\centering%
\begin{tabular}{cc}
\scalebox{0.34}{\includegraphics{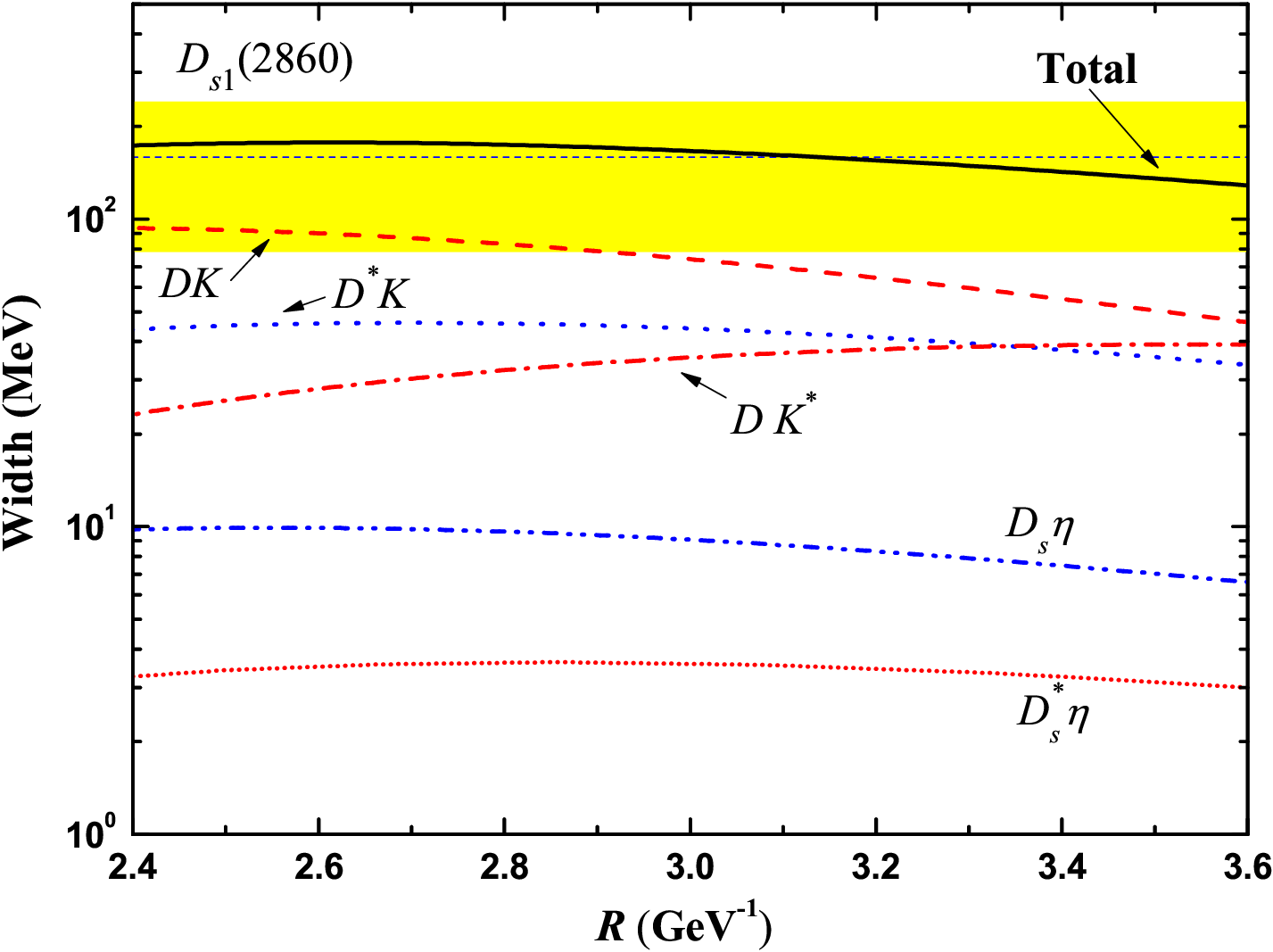}}&%
\scalebox{0.34}{\includegraphics{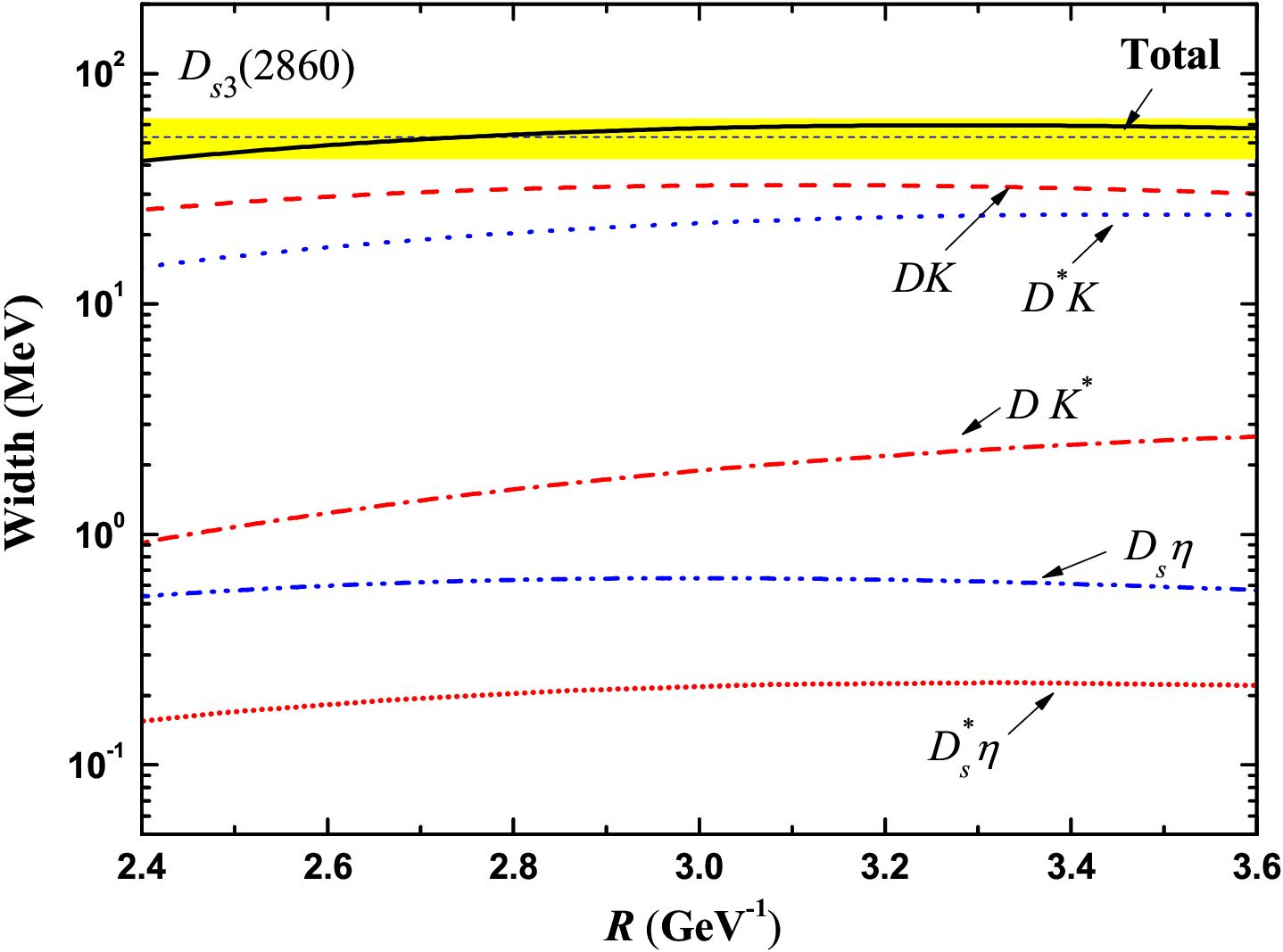}}%
\end{tabular}
\caption{(Color online). The total and partial decay widths of
$1^3D_1$ (left panel) and $1^3D_3$ (right panel) charmed-strange mesons
dependent on the $R$ value.
Here, the dashed lines with the yellow bands are the experimental
widths of $D_{s1}^*(2860)$ and $D_{s3}^*(2860)$ from LHCb
\cite{Aaij:2014xza,Aaij:2014baa}. \label{Fig:3D1D3}}
\end{figure*}

With the above preparation, we obtain the total and partial decay
widths of $D_{s1}^*(2860)$, $D_{s3}^*(2860)$ and their spin
partners, and comparison with the experimental data
\cite{Aaij:2014xza,Aaij:2014baa}. As shown in Table
\ref{table:Rvalue}, the $R$ value of the $P$-wave charmed-strange meson
is about 2.70 $\mathrm{GeV}^{-1}$. For the $D$-wave state, the $R$ value
is estimated to be around 3.00 $\mathrm{GeV}^{-1}$
\cite{Close:2005se}. In present calculation, we vary the $R$ value
for the $D$-wave charmed-strange meson from 2.4 to 3.6
$\mathrm{GeV}^{-1}$. In Fig. \ref{Fig:3D1D3}, we present the $R$ dependence of the
total and partial decay widths of $D_{s1}(2860)$ and
$D_{s3}(2860)$.

\subsection{$D_{s1}(2860)$}

The total width of $D_{s1}(2860)$ as the $1^3D_1$ state is given in Fig. \ref{Fig:3D1D3}, where
the total width is obtained as $128\sim$ 177 MeV corresponding to the selected $R$ range, which is consistent with the experimental width of $D_{s1}(2860)$ ($\Gamma=159 \pm 23 \pm 27\pm
72$ MeV \cite{Aaij:2014xza,Aaij:2014baa}). The information of the partial decay widths depicted in Fig. \ref{Fig:3D1D3} also shows that $DK$ is the dominant decay mode of the $1^3D_1$
charm strange meson, which explains why $D_{s1}(2860)$ was experimentally observed in its $DK$ decay channel \cite{Aaij:2014xza,Aaij:2014baa}). In addition, our study also
indicates that the $D^\ast K$ and $D
K^\ast$ channels are also important for the $1^3D_1$ state, which have partial widths $35 \sim 44$ MeV and
$24 \sim 40$ MeV, respectively. The $D_s \eta$ and $D_s^\ast \eta$ channels have partial decay widths with several MeV, which is far smaller than the partial decay widths of the $DK$, $D^\ast K$ and $D K^\ast$ channels. This phenomena can be understood since the decays of the $1^3D_1$ state into
$D_s \eta$ and $D_s^\ast \eta$ have the smaller phase space.

The above results show that $D_{s1}(2860)$
can be a good candidate for the $1^3D_1$ charmed-strange meson.

Besides providing the partial decay widths, we also predict several typical ratios, i.e.,
\begin{eqnarray}
\frac{\Gamma(D_{s1}(2860) \to D^\ast K)}{\Gamma(D_{s1}(2860) \to D
K)}&=& 0.46 \sim 0.70,\label{aa1}\\
\frac{\Gamma(D_{s1}(2860) \to D K^\ast)}{\Gamma(D_{s1}(2860) \to D
K)} &=& 0.25 \sim 0.85,\\
\frac{\Gamma(D_{s1}(2860) \to D_s \eta)}{\Gamma(D_{s1}(2860) \to D
K)} &=& 0.10 \sim 0.14,
\end{eqnarray}
which can be further tested in the coming experimental
measurements.

{The Belle Collaboration once reported $D_{s1}(2710)$ in the $DK$ invariant mass spectrum, which has mass $2708\pm9^{+11}_{-10}$ MeV and width $108\pm23^{+36}_{-31}$ MeV \cite{Brodzicka:2007aa}. The analysis of angular distribution indicates that
$D_{s1}(2710)$ has the spin-parity $J^P=1^-$ \cite{Brodzicka:2007aa}. Later, the BaBar Collaboration confirmed $D_{s1}(2710)$ in a new $D^*K$ channel \cite{Aubert:2009ah}, where the ratio $\Gamma(D^*K)/\Gamma(DK)=0.91\pm0.13\pm0.12$ for $D_{s1}(2710)$.

In Refs. \cite{Zhang:2006yj,Colangelo:2007ds},
the assignment of $D_{s1}(2710)$ to the $1^3D_1$ charmed-strange meson was proposed. However, the obtained $\Gamma(D^*K)/\Gamma(DK)=0.043$ \cite{Colangelo:2007ds} is deviated far from the experimental data, which does not support  the $1^3D_1$ charmed-strange assignment to $D_{s1}(2710)$. Especially, the present study of newly observed $D_{s1}(2860)$ shows that $D_{s1}(2860)$ is a good candidate of the $1^3D_1$ charmed-strange meson.}

{If $D_{s1}(2710)$ is not a $1^3D_1$ charmed-strange meson, we need to find the place in the charmed-strange meson family. In Ref. \cite{Zhang:2006yj}, the authors once indicated that $D_{s1}(2710)$ as the $2^3S_1$ charmed-strange meson is not completely excluded\footnote{We need to explain the reason why the $2^3S_1$ charmed-strange meson is not completely excluded in Ref. \cite{Zhang:2006yj}. In Ref. \cite{Zhang:2006yj}, the total decay width of $D_{s1}(2710)$ as the $2^3S_1$ charmed-strange meson was calculated, which is 32 MeV. This value is obtained with the typical $R=3.2$ GeV$^{-1}$ value. As shown in Fig. 4 (d) in \cite{Zhang:2006yj}, the total decay width is strongly dependent on the $R$ value due to node effects. If taking other typical $R$ values which are not far away from  $3.2$ GeV$^{-1}$, the total decay width can reach up to the lower limit of the experimental width of $D_{s1}(2710)$.}. By the effective Lagrangian approach \cite{Colangelo:2007ds} and under
the assignment of $D_{s1}(2710)$ to the $2^3S_1$ charmed-strange meson,
the $\Gamma(D_{s1}(2710)\to D^*K)/\Gamma(D_{s1}(2710)\to DK)=0.91$ was obtained, which is close to the experimental value \cite{Brodzicka:2007aa}. We also notice the recent work of $D_{sJ}(2860)$ \cite{Godfrey:2014fga}, where they also proposed that $D_{s1}(2710)$ is the $2^3S_1$ $c\bar{s}$ meson. }

{Besides the above exploration in the framework of a conventional charmed-strange meson, the exotic state explanation to $D_{s1}(2710)$ was given in Ref. \cite{Vijande:2008zn}.}

\begin{figure*}[htb]
\centering%
\scalebox{1}{\includegraphics{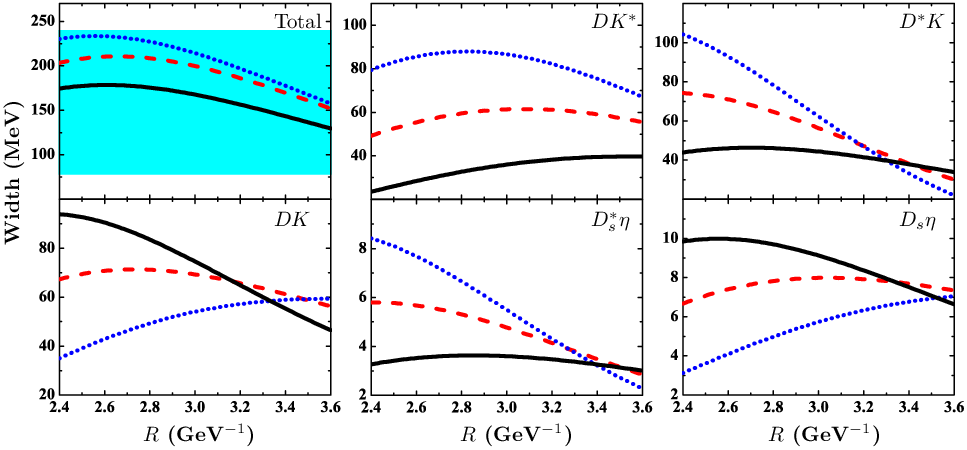}}  %
\caption{(Color online). {The total and partial decay widths (in units of MeV) of $D_{s1}(2860)$ dependent on the $R$ value. The solid, dashed and dotted curves correspond to the typical $2S$-$1D$ mixing angles $\theta= 0^\circ $, $\theta =15^\circ$ and $\theta=30^\circ$, respectively. The band in the left-top panel is the total decay width with errors, which was reported by the LHCb Collaboration \cite{Aaij:2014baa, Aaij:2014xza}. }\label{Fig:SDMixing}}
\end{figure*}

In the following, we include the mixing between $2^3S_1$ and $1^3D_1$ states to further discuss the mixing angle dependence of the decay behavior of $D_{s1}(2860)$. Here, $D_{s1}(2S)$ and $D_{s1}(2860)$ are the mixtures of $2^3S_1$ and $1^3D_1$ states, which satisfy the relation below
\begin{eqnarray}
\left(
\begin{array}{c}
  |D_{s1}(2S)\rangle \\
  |D_{s1}(2860)\rangle
\end{array}
\right) =
\left(
\begin{array}{cc}
 \cos \theta  & \sin \theta \\
-\sin \theta  & \cos \theta
\end{array}
\right)
\left(
\begin{array}{c}
  |2^3S_1\rangle\\
  |1^3D_1 \rangle
\end{array}
\right).
\end{eqnarray}
{
The $2S$-$1D$ mixing angle should be small due to the relative large mass gap between $2S$ and $1D$ state, where we take some typical  values $\theta= 0^\circ$, $\theta=15^\circ$, and $\theta=30^\circ$ to present our results. $\theta=0^\circ$ denotes that there is no $2S$-$1D$ mixing. In Fig. \ref{Fig:SDMixing}, we present the total and partial decay widths, which depend on the $R$ value, where we show the results with three different typical $\theta$ values. In the left-top panel of Fig. \ref{Fig:SDMixing}, we list the total decay width of $D_{s1}(2860)$ and the comparison with the experimental data. When taking $2.4\ \mathrm{GeV}^{-1}<R<3.6\ \mathrm{GeV}^{-1}$, the calculated total decay width varies from 130 MeV to 235 MeV, which indicates the theoretically estimated total decay width of $D_{s1}(2860)$ can overlap with the experimental measurement. In addition, we also notice that the partial decay widths for the $ D_s \eta$ and $ D_s^\ast \eta $ channels are less than 10 MeV, while for the $DK$, $D^\ast K$ and $D K^\ast$ decay modes, the corresponding partial decay widths vary from several 20 MeV to more than one hundred MeV, which strongly depend on the value of a mixing angle.

In Ref. \cite{Li:2009qu}, the authors adopted a different convention for the $2S$-$1D$ mixing, where their $2S$-$1D$ mixing angle has a sign opposite to our scenario. Taking the same $R$ value and mixing angle for $D_{s1}(2860)$, we obtain the partial decay widths of $DK$ and $D^\ast K$ are less than 10 MeV, while the calculated partial decay width of $DK^\ast$ is more than 20 MeV, which is consistent with the results in Ref. \cite{Li:2009qu}.
}

\subsection{$D_{s3}(2860)$}

The two-body OZI-allowed decay behavior of $D_{s3}(2860)$ as the $1^3D_3$ charmed-strange meson is presented in the right panel of Fig. \ref{Fig:3D1D3}, where the obtained total width can reach up to $42 \sim 60$ MeV, which
overlaps with the LHCb's data ($53 \pm 7 \pm 4 \pm 6$
MeV \cite{Aaij:2014xza,Aaij:2014baa}). This fact further reflects that $D_{s3}(2860)$ is suitable for the $1^3D_3$ charmed-strange meson.
Similar to $D_{s1}(2860)$, the $DK$ channel is also the dominant decay mode of
$D_{s3}(2860)$, where the partial decay width of $D_{s3}(2860) \to DK$
is $25 \sim 30$ MeV in the selected $R$ value range. Additionally, we also calculate
the partial decay width of $D_{s3}(2860) \to D^\ast K$
and $D_{s3}(2860) \to D K^\ast $, which are $14 \sim 24$ MeV and $0.9 \sim
2.5$ MeV, respectively. The partial decay widths for $D_s \eta$ and
$D_s^\ast \eta$ channel are of order of 0.1 MeV. The corresponding typical
ratios for $D_{s3}(2860)$ are
\begin{eqnarray}
\frac{\Gamma(D_{s3}(2860) \to D^\ast K)}{\Gamma(D_{s3}(2860) \to D
K)} &=& 0.55 \sim 0.80,\label{bb1}\\
\frac{\Gamma(D_{s3}(2860) \to D K^\ast)}{\Gamma(D_{s3}(2860) \to D
K)} &=& 0.03 \sim 0.09,\\
\frac{\Gamma(D_{s3}(2860) \to D_s \eta)}{\Gamma(D_{s1}(2860) \to D
K)} &=& 0.018 \sim 0.020,
\end{eqnarray}
which can be tested in future experiment.

{In Ref. \cite{Zhang:2006yj}, the ratio $\Gamma(D^*K)/\Gamma(DK)=13/22=0.59$ was given for $D_{sJ}(2860)$ observed by Belle as the $1^3D_3$ state, where the QPC model was also adopted and this ratio is obtained by taking a typical value $R=2.94$ GeV$^{-1}$. In the present work, we consider the range $R=2.4\sim 3.6$ GeV$^{-1}$ to present the $D^*K/DK$ ratio for $D_{s3}(2860)$. If comparing the ratio given in Eq. (\ref{bb1}) with the corresponding one obtained in Ref. \cite{Zhang:2006yj}, we notice that their value $\Gamma(D^*K)/\Gamma(DK)=0.59$ \cite{Zhang:2006yj} just falls into our obtained range $\Gamma(D^*K)/\Gamma(DK)=0.55\sim 0.80$. }

{In addition, we need to make a comment on the experimental ratio $\Gamma(D^*K)/\Gamma(DK)=1.10\pm0.15\pm0.19$ for $D_{sJ}(2860)$, which was extracted by the BaBar Collaboration \cite{Aubert:2006mh}. Since LHCb indicated that there exist two resonances $D_{s1}(2860)$ and $D_{s3}(2860)$ in the $DK$ invariant mass spectrum \cite{Aaij:2014xza,Aaij:2014baa}, the experimental ratio $\Gamma(D^*K)/\Gamma(DK)$ for $D_{sJ}(2860)$ must be changed, which means that we cannot simply apply the old $\Gamma(D^*K)/\Gamma(DK)$ data in Ref. \cite{Aubert:2006mh} to draw a conclusion on the structure of $D_{sJ}(2860)$.
We expect further experimental measurements of this ratio, which will be helpful to reveal the properties of the observed $D_{sJ}(2860)$ states. }

\begin{figure*}[htbp]
\centering%
\scalebox{1}{\includegraphics{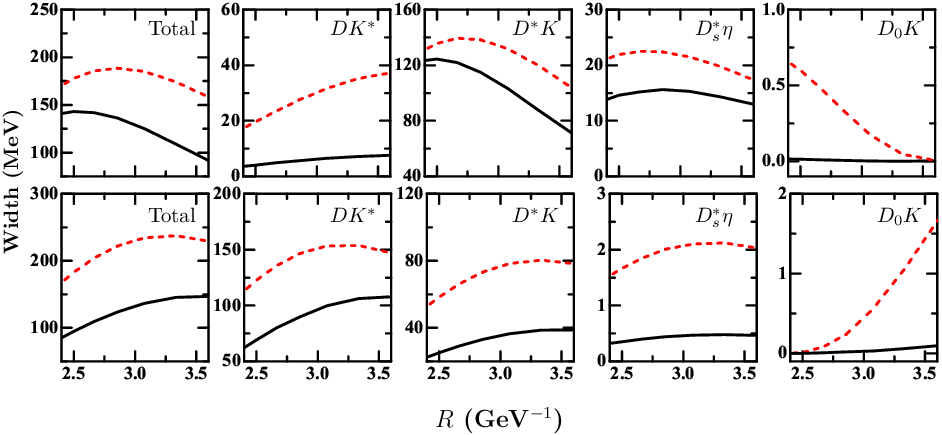}}  %
\caption{(Color online). {The total and partial decay widths of
$1D(2^-)$ (upper row) and $1D^\prime(2^-)$ (lower row)
charmed-strange mesons dependent on the $R$ values. Here, a mixing angle of $39^\circ$ is chosen. The solid and dashed curves correspond to two predicted masses of $2^-$ states, 2850 MeV and 2950 MeV, respectively. }\label{Fig:D2}}
\end{figure*}

\subsection{$1D(2^-)$ and $1D^\prime(2^-)$}

In the following, we discuss
the decay behaviors of two missing $1D$ states in the present
experiment, which is crucial to the experimental search for the
$1D(2^-)$ and $1D^\prime(2^-)$ states.

{
We first fix the mixing angle $\theta_{1D}=39^\circ$ obtained in the heavy quark limit, and discuss the $R$ value dependence of the total and partial decay widths of the two missing $2^-$ states in experiment, which are presented in Fig. \ref{Fig:D2}.} Since these two $1D$ states have not yet been seen
in experiment, we take the mass range $2850 \sim 2950$ MeV, which covers the theoretically predicted masses of the $1D(2^-)$ and $1D^\prime(2^-)$ states from different groups listed in Table \ref{prediction}, to discuss the decay behaviors of the $1D(2^-)$ and $1D^\prime(2^-)$ states.

The total and partial decay width of $1D(2^-)$ are present in the upper panel of Fig. \ref{Fig:D2}. Here, two typical mass of $1D(2^-)$ state, 2850 MeV and 2950 MeV, are considered, which are corresponding to the solid and dashed curved in Fig. \ref{Fig:D2}. The estimated total decay width varies from 90 MeV to 190 MeV, which is due to the uncertainty of the predicted mass of the $1D(2^-)$ state and the $R$ value dependence as mentioned above. If the mass of the $1D(2^-)$ state can be constrained by future experiment, the uncertainty of the total width for the $1D(2^-)$ state can be further reduced. In any case, our study indicates that the $1D(2^-)$ state has a broad width.

Additionally, as shown in Fig. \ref{Fig:D2}, the $1D(2^-)$ state can dominantly decay into $D^\ast K$, where $\mathcal{B}(1D(2^-)\to D^\ast K)=(77\sim 87)\%$, and $D K^\ast$ and
$D_s^\ast \eta$ are its main decay modes. Comparing $D^\ast K$, $DK^\ast$ and $D_s^\ast \eta$ with each other, $D_s^\ast \eta$ is the weakest decay channel. Hence, we suggest experimental search for the $1D(2^-)$ state firstly via the $D^\ast K$ channel.

We also obtain two typical ratios, i.e.,
\begin{eqnarray}
&&\frac{\Gamma(1D(2^-)\to D K^\ast)}{\Gamma(1D(2^-)\to D^\ast K) }
=0.11\sim 0.36\\
&&\frac{\Gamma(1D(2^-)\to D_s^\ast \eta)}{\Gamma(1D(2^-)\to D^\ast
K)
} =0.11\sim 0.18, \label{Eq:1D2b}
\end{eqnarray}
which can be accessible in experiment.

As for the $1D^\prime(2^-)$ state, the total and partial decay width are present in the lower panel of Fig. \ref{Fig:D2}. we predict its total decay width
($(80\sim240)$ MeV), which shows that the $1D^\prime(2^-)$ state is also a broad resonance, where $DK^\ast$ is its dominant
decay mode with a branching ratio $\mathcal{B}(1D^\prime (2^-)\to D K^\ast)=(0.64\sim 0.73)\%$. Its main decay mode includes $D^\ast K$, while $1D^\prime \to D_s^\ast
\eta$ and $1D^\prime \to D_0(2400) K$ have small partial decay widths. Besides the above information, two typical
ratios are listed as below
\begin{eqnarray}
&&\frac{\Gamma(1D^\prime (2^-)\to D^\ast K)}{\Gamma(1D^\prime
(2^-)\to D K^\ast)} =0.36 \sim 0.53,\\
&&\frac{\Gamma(1D^\prime (2^-)\to D^\ast \eta)}{\Gamma(1D^\prime
(2^-)\to D K^\ast)} =0.004 \sim 0.013.
\end{eqnarray}

\renewcommand{\arraystretch}{1.5}
\begin{table*}[htbp]
\caption{{The total and partial decay widths of two $2^-$ states (in units of MeV) dependent on the mixing angle $\theta_{1D}$ (The typical values are $\theta_{1D}=20^\circ,\ 30^\circ$ and $50^\circ$). Here, the $R$ value of $1D(2^-)$ and $1D'(2^-)$ is fixed as $R=2.85$ GeV$^{-1}$ \cite{Close:2005se}.}\label{Tab:D2} }
\centering
\begin{tabular}{l c c c c c c| c c c c c c  }\toprule[1pt]

&\multicolumn{6}{c|}{M=2850 MeV}&\multicolumn{6}{c}{M=2950 MeV}\\
&\multicolumn{2}{c}{$\theta_{1D}=20^\circ$}  &\multicolumn{2}{c}{$\theta_{1D}=30^\circ$} & \multicolumn{2}{c|}{$\theta_{1D}=50^\circ$} & \multicolumn{2}{c}{$\theta_{1D}=20^\circ$} & \multicolumn{2}{c}{$\theta_{1D}=30^\circ$} & \multicolumn{2}{c}{$\theta_{1D}=50^\circ$} \\
Channels&$ 1D(2^-)$&$ 1D'(2^-) $&$ 1D(2^-)$&$ 1D'(2^-) $&$ 1D(2^-)$&$ 1D'(2^-) $&$ 1D(2^-)$&$ 1D'(2^-) $&$ 1D(2^-)$&$ 1D'(2^-) $&$ 1D(2^-)$&$ 1D'(2^-) $ \\ \midrule[1pt]
$D^*K$&126.93&44.60&135.61&35.90&134.62&36.84&110.42&77.82&113.85 &74.39&113.46&74.76 \\
$DK^*$&26.06&70.04&13.54&82.59&1.99&94.18&56.19&118.06&38.58&135.69 &22.34&152.00 \\
$D_s^* \eta$&13.93&2.08&15.17&0.83&15.03&0.97&20.23&4.21&21.91&2.54 &21.72&2.72 \\
$D_0^*(2400)K$&0.012&0.011&0.008&0.015&0.002&0.022&0.50&0.08&0.42&0.16  &0.22&0.36\\ \midrule[1pt]
Total& 166.93&116.73&164.33&119.34 &151.64 &132.01 &187.34 & 200.17& 174.76& 212.78 &157.74&229.84 \\ \bottomrule[1pt]
\end {tabular}\\
\label{table:angle} 
\end{table*}

It should be noticed that the threshold of $D_{s0}^\ast (2317) \eta$
is 2865 MeV and two $1D$ charmed-strange mesons with $J^P=2^-$ decaying
into $D_{s0}^\ast(2317) \eta$ occur via $D-$wave. Thus, $1D(2^-)/1D^\prime(2^-)\to D_{s0}^\ast (2317) \eta$ is suppressed, which is supported by our calculation since the corresponding partial decay widths are
of the order of a few keV for $1D(2^-) \to D_{s0}^\ast \eta$ and of the order of 0.1 keV
for $1D^\prime(2^-) \to D_{s0}^\ast \eta$.

{ In Table \ref{Tab:D2}, we fix the $R$ value of $1D(2^-)$ and $1D^\prime(2^-)$ to be 2.85 GeV$^{-1}$ \cite{Close:2005se} and further discuss the total and partial decay widths dependent on the mixing angle $\theta_{1D}$, where three typical values $\theta_{1D}=20^\circ$, $\theta_{1D}=30^\circ$ and $\theta_{1D} =50^\circ$ are adopted. For $1D(2^-)$ state, its total decay width varies from 152 MeV to 187 MeV, which indicates that the total decay width is weakly dependent on the mixing angle $\theta_{1D}$. Moreover, $1D(2^-)$ state dominantly decays into $D^\ast K$, whose width varies from 110 MeV to 134 MeV caused by the uncertainty of the mass of $1D(2^-)$ state and the different mixing angle $\theta_{1D}$. In addition,  the ratio of $1D(2^-) \to D_s^\ast \eta$ and $1D(2^-) \to D^\ast K$ is estimated to be $0.11 \sim 0.19$, which is consistent with that shown in Eq. (\ref{Eq:1D2b}). However, the predicted partial decay width of $1D(2^-) \to D K^\ast$ is strongly dependent on the mixing angle, which varies from 2 MeV to 56 MeV.

The total decay width of $1D^\prime(2^-)$ varies from 117 MeV to 230 MeV depending on different predicted masses and mixing angles. Its dominant decay modes are $D^\ast K $ and $DK^\ast$ and the ratios of $1D^\prime(2^-) \to D^\ast K$ and $1D^\prime(2^-) \to D K^\ast$ are predicted to be $0.39 \sim 0.66$. The partial decay widths of $1D^\prime(2^-) \to D_s^\ast \eta$ and $1D^\prime(2^-) \to D_0(2400) K$ are relatively small and are given by several MeV and less than 0.5 MeV, respectively.
}

\section{Discussion and conclusions}\label{sec3}

With the observation of two charmed-strange
resonances $D_{s1}(2860)$ and  $D_{s3}(2860)$, which was recently announced by
the LHCb Collaboration \cite{Aaij:2014xza,Aaij:2014baa},
the observed charmed-strange states become more and more abundant.
In this work, we have carried out a study of the observed $D_{s1}(2860)$
and  $D_{s3}(2860)$, which indicates that $D_{s1}(2860)$ and
$D_{s3}(2860)$ can be well categorized as $1^3D_1$ and $1^3D_3$ states in
the charmed-strange meson family since the experimental widths
of $D_{s1}(2860)$ and  $D_{s3}(2860)$ can be reproduced by the
corresponding calculated total widths of the $1^3D_1$ and $1^3D_3$
states. In addition, the result of their partial decay widths shows
that the $DK$ decay mode is dominant both for $1^3D_1$ and
$1^3D_3$ states, which naturally explains why $D_{s1}(2860)$ and
$D_{s3}(2860)$ were first observed in the $DK$ channel. If
$D_{s1}(2860)$ and  $D_{s3}(2860)$ are the $1^3D_1$ and $1^3D_3$
states, respectively, our study also indicates that the $D^*K$ and
$DK^*$ channels are the main decay mode of $D_{s1}(2860)$ and
$D_{s3}(2860)$, respectively. Thus, we suggest for future experiment to search for
$D_{s1}(2860)$ and  $D_{s3}(2860)$ in these main decay channels,
which can not only test our prediction presented in this work but
also provide more information of the properties of $D_{s1}(2860)$
and  $D_{s3}(2860)$.

As the spin partners of $D_{s1}(2860)$ and  $D_{s3}(2860)$, two $1D$
charmed-strange mesons with $J^P=2^-$ are still missing in
experiment. Thus, in this work we also predict the decay behaviors of
these two missing $1D$ charmed-strange mesons. Our calculation by
the QPC model shows that both $1D(2^-)$ and $1D^\prime(2^-)$ states
have very broad widths. For the $1D(2^-)$ and $1D^\prime(2^-)$
states, their dominant decay mode is $D^*K$ and $DK^*$, respectively. In addition,
$DK^*$ and $D^*K$ are also the important decay modes of the $1D(2^-)$ and
$1D^\prime(2^-)$ states, respectively. This investigation provides
valuable information when further experimentally exploring these two
missing $1D$ charmed-strange mesons.

In summary, the observed  $D_{s1}(2860)$ and  $D_{s3}(2860)$ provide
us a good opportunity to establish higher states in the
charmed-strange meson family. The following experimental and
theoretical efforts are still necessary to reveal the underlying
properties of  $D_{s1}(2860)$ and  $D_{s3}(2860)$. Furthermore, it
is a challenging research topic for future experiment to hunt
the two predicted missing $1D$ charmed-strange mesons with
$J^P=2^-$.

Before closing this section, we would like to discuss the threshold effect or coupled-channel effect, which was proposed to
solve the puzzling lower mass and narrow widths for $D_{s0}(2317)$ \cite{vanBeveren:2003kd} and $D_{s1}(2460)$ \cite{vanBeveren:2003jv}, and to understand the properties of $X(3872)$. We notice that there exist several typical $D^*K$, $DK^*$, and $D^*K^*$ thresholds, which are $\sim2580$ MeV, $\sim2762$ MeV, and $\sim 2902$ MeV, respectively. Here, the observed $D_{s1}(2860)$ and  $D_{s3}(2860)$ are near the $D^*K^*$ threshold while $D_{s1}(2715)$ is close to the $DK^*$ threshold. This fact also shows that the threshold effect or coupled-channel effect is important to these observed charmed-strange states. For example, in Ref. \cite{Molina:2010tx}, the authors studied the $D^*K^*$ threshold effect on $D_{s2}^*(2573)$. Thus, further theoretical study of $D_{s1}(2860)$ and  $D_{s3}(2860)$ by considering the threshold effect or coupled-channel effect is an interesting research topic.

{In addition, the results presented in this work are calculated by using the SHO wave functions with a rms radius obtained within a relativistic quark model \cite{Godfrey:1985xj}, which can provide a quantitative estimate of the decay behavior of hadrons.  However, the line shape of the SHO wave function is slightly different from the one obtained by the relativistic quark model. For example, the nodes may appear at different places for these two cases. Thus,
adopting a numerical wave function from a relativistic quark model \cite{Godfrey:1985xj} may further improve the whole results, where we can compare the results obtained by using the SHO wave function and the numerical wave function, which
 is an interesting research topic\footnote{We would like to thank the anonymous referee for his/her valuable suggestion.}. }


\section*{Acknowledgments}

This project is supported by the National Natural Science Foundation
of China under Grants No. 11222547, No. 11375240, No. 11175073, and
No. 11035006, the Ministry of Education of China (SRFDP under Grant
No. 2012021111000), and the Fok Ying Tung Education Foundation (No.
131006).

\end{document}